\documentstyle[twoside,fleqn,espcrc2,epsf]{article}

\newcommand{\lw}[1]{\smash{\lower2.ex\hbox{#1}}}
\def\simlt{\rlap{\lower 3.5 pt\hbox{$\mathchar \sim$}}\raise 1pt \hbox {$<$}}
\def\simgt{\rlap{\lower 3.5 pt\hbox{$\mathchar \sim$}}\raise 1pt \hbox {$>$}}

\newcommand{\AmS}{{\protect\the\textfont2
  A\kern-.1667em\lower.5ex\hbox{M}\kern-.125emS}}

\title{
\vspace{-2.8cm}
\begin{flushleft}
       {\normalsize UTCCP-P-50,\ UTHEP-391}   \\[-0.2cm]
       {\normalsize September 1998}   \\
\end{flushleft}
       \vspace{0.7cm}
Quenched Light Hadron Spectrum with the Wilson Quark Action:\\
Final Results from CP-PACS
\thanks{Presented by T. Yoshi\'e
at ``Lattice 98'', Boulder, Colorado, USA, 13--18 July 1998.}}

\author{CP-PACS Collaboration :\\
        S.~Aoki\rlap,\address{Institute of Physics, University of
        Tsukuba, Tsukuba, Ibaraki 305-8571, Japan}
        G.~Boyd\rlap,\address{Center for Computational Physics,
        University of Tsukuba, Tsukuba, Ibaraki 305-8577, Japan}
        R.~Burkhalter\rlap,$^{\rm a,b}$
        S.~Ejiri\rlap,$^{\rm b}$
        M.~Fukugita\rlap,\address{Institute for Cosmic Ray Research,
        University of Tokyo, Tanashi, Tokyo 188-8502, Japan}
        S.~Hashimoto\rlap,\address{High Energy Accelerator Research Organization
        (KEK), Tsukuba, Ibaraki 305-0801, Japan}
        Y.~Iwasaki\rlap,$^{\rm a,b}$
        K.~Kanaya\rlap,$^{\rm a,b}$
        T.~Kaneko\rlap,$^{\rm b}$
        Y.~Kuramashi\rlap,$^{\rm d}$
        K.~Nagai\rlap,$^{\rm b}$
        M.~Okawa\rlap,$^{\rm d}$
        H.P.~Shanahan\rlap,$^{\rm b}$
        A.~Ukawa\rlap,$^{\rm a,b}$ and
        T.~Yoshi\'e$^{\rm a,b}$ }

\begin{document}

\begin{abstract}
We report the final results of the CP-PACS calculation for
the quenched light hadron spectrum with the Wilson quark action.
Our data support the presence of quenched chiral singularities,
and this motivates us to use
mass formulae based on quenched chiral perturbation theory
in order to extrapolate hadron masses to the physical point.
Hadron masses and decay constants in the continuum limit 
show unambiguous systematic deviations from experiment.
We also report the results for light quark masses. 
\end{abstract}

\maketitle

\section{Introduction}
At Lattice'97 we presented first results from 
the CP-PACS calculation of the quenched light hadron spectrum with 
the Wilson quark action on large lattices ($La \simgt 3$ fm) 
at small quark masses ($m_\pi/m_\rho=$ 0.75 down to 0.4) 
with high statistics
(800, 600, 420 and 91 configurations at $\beta=$ 5.9, 6.1, 6.25 and 6.47)
\cite{ref:CPPACSlat97}. 
We have since increased the statistics at $\beta=6.47$ to 150, and 
have completed the analysis.  
In this article, we report the final spectrum results and the 
main points of analyses behind them. 

\section{Quenched chiral singularities}
Chiral extrapolation is a basic element of spectrum calculations, for 
which a choice has to be made of the functional form of hadron masses
in terms of quark masses.
An important issue in considering the choice is the validity of 
quenched chiral perturbation theory(Q$\chi$PT) \cite{ref:QChPTSharpe,ref:QChPTBGmassratio,ref:Booth,ref:LabrenzSharpe}, which predicts 
characteristic singularities in hadron masses in the chiral limit. 
We have therefore made a detailed examination of this issue. 

\subsection{pseudo-scalar mesons}
For pseudo-scalar (PS) mesons made of quarks of mass 
$m$ and $m_s$, Q$\chi$PT formula reads\cite{ref:QChPTBGmassratio} 
\begin{eqnarray}
m_{PS}^2 &=&  A(m_s+m)\{ 1 - \delta [\ln(2mA/\Lambda_\chi^2) \nonumber \\
  &+& m_s/(m_s-m) \ln(m_s/m) ] \} \nonumber \\
  &+& B(m_s+m)^2  + C(m_s-m)^2 + \cdots . \label{eq:mps}
\end{eqnarray}
To test the presence of the logarithm term, we combine our results to 
form two quantities  
\begin{eqnarray}
y &=& {\frac{2m}{m_s+m} \frac{m_K^2}{m_\pi^2}} \times 
   {\frac{2m_s} {m_s+m} \frac{m_K^2}{m_\eta^2}}, \\
x &=& 2 - \frac{m_s+m}{m_s-m}\log(\frac{m_s}{m}),\label{eq:ratiox}
\end{eqnarray}
where $\pi\ (\eta)$ is the degenerate PS meson with 
quark mass $m\ (m_s)$ at $m_\pi/m_\rho=$ 0.6, 0.5, 0.4\ (0.75, 0.7) 
and $K$ is the non-degenerate one with $m$ and $m_s$. 
The two quantities are related by 
$y = 1 + \delta\cdot x$, where the leading correction depends only 
on the $O((m_s-m)^2)$ term in (\ref{eq:mps}).

\begin{figure}[t]
\begin{center} \leavevmode
\epsfxsize=7.0cm \epsfbox{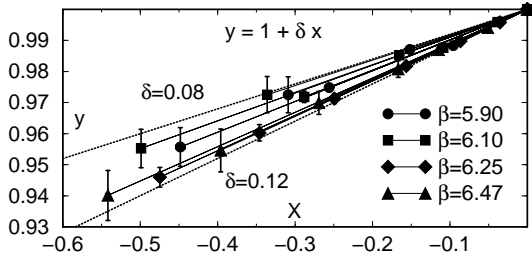}
\end{center}
\vspace{-15mm}
\caption{Results of a test for the presence of 
the quenched chiral logarithms in PS meson mass data. 
See text for details.}
\label{fig:PSMassRatio}
\vspace{-7mm}
\end{figure}

In Fig.~\ref{fig:PSMassRatio} we plot the two quantities
calculated with quark masses determined from an extended axial current Ward
identity ($m_q^{AWI}$) as they have no ambiguity associated with the 
determination of the critical hopping parameter. 
The data fall within a narrow wedge spanned by the lines 
$y \approx 1 + (0.08$ -- $0.12) x$, implying the value $\delta \approx 0.10(2)$.
We note that the $O((m_s+m)^2)$ term can not be ignored 
for the range of our quark masses; 
results for the original ratio 
$m_K^2/m_\pi^2$\cite{ref:QChPTBGmassratio}, which receive corrections 
both from $O((m_s+m)^2)$ and $O((m_s-m)^2)$ terms, 
do not fall on a common line. 

A different test using a ratio of decay constants 
$y = f_K^2/(f_\pi f_\eta)$\cite{ref:QChPTBGmassratio} leads to
a similar result; our data fall within the lines 
$y=1- \delta/2 \cdot x$ with $\delta=0.08$ -- 0.16.

Finally, making full correlated fits to $m_{PS}$ using (\ref{eq:mps})
but imposing $C=0$, independently for degenerate 
and non-degenerate data, we find $\delta \approx 0.06$ -- 0.12 
for the range 
$\Lambda_\chi \approx 0.6$ -- 1.4 GeV.

These results lead us to conclude that our PS data show evidence for 
Q$\chi$PT logarithms. 

\subsection{vector meson and baryon masses}

For vector mesons and baryons, we perform uncorrelated simultaneous fits
to degenerate and non-degenerate data together as a function of $m_{PS}$,
assuming Q$\chi$PT mass formulae\cite{ref:Booth,ref:LabrenzSharpe} 
with $\delta=0.1$.
For vector mesons and decuplet baryons,
all $O(m_{PS})$ terms of Q$\chi$PT are included as well as 
$O(m_{PS}^2)$ terms.
For octet baryons, we include $O(m_{PS}^3)$ terms 
in addition to $O(m_{PS})$ and $O(m_{PS}^2)$ terms
since the nucleon mass shows a negative curvature which is 
opposite to that of the $O(m_{PS})$ term.
We omit decuplet-octet coupling ($C$ in the notation of 
Ref.~\cite{ref:LabrenzSharpe}) and coupling to $\eta'$ ($\gamma$), 
and set $\alpha_\Phi=0$. 

These mass formulae fit our data well. 
The values for the coefficient $C_{1/2}$ of $O(m_{PS})$ terms, 
however, are small. 
We obtain $C_{1/2}$ = $-0.071(8)$ for $\rho$,
$-0.118(4)$ for nucleon, and $-0.14(1)$ for $\Delta$, 
to be compared with phenomenological estimates: 
$C_{1/2}=- 4\pi g_2^2 \delta \approx -0.71$ for $\rho$ 
if $\delta=0.1$ and $g_2=0.75$, and 
$C_{1/2} = -(3\pi/2) (D-3F)^2\delta \approx -0.27$ for nucleon 
if $\delta=0.1$ and $(F,D)=(0.5,0.75)$. 

We conclude that vector and baryon masses are compatible with 
the presence of $O(m_{PS})$ terms, 
but that their magnitudes are smaller than expected.
These terms yield an effect of at most 
$0.2\times m_\pi \approx$ 25 MeV, which is about 3\% of light hadron masses.

\section{Final spectrum results}
\vspace{-2mm}
\begin{figure}[t]
\begin{center} \leavevmode
\epsfxsize=7.5cm \epsfbox{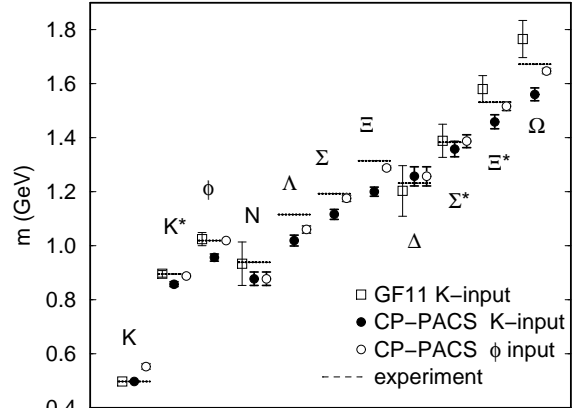}
\end{center}
\vspace{-15mm}
\caption{Final spectrum results compared to
GF11's\protect\cite{ref:GF11mass} and experiment.}
\label{fig:spectrum}
\vspace{-7mm}
\end{figure}

The results of analyses above motivate us to adopt the Q$\chi$PT fits 
to calculate masses at each value of $\beta$. 
The physical point for degenerate $u$ and $d$ quarks and the lattice scale are 
determined from the experimental values of $m_\pi$ and $m_\rho$, 
and the strange quark mass by that of $m_K$ or $m_\phi$.
We then extrapolate the results linearly in $a$.
The final result for the spectrum in the continuum limit is shown 
in Fig.\ref{fig:spectrum}.
	
In order to examine how results differ if we do not employ 
Q$\chi$PT mass formulae, 
we repeat the analysis employing a quadratic polynomial in $1/K$ 
(cubic for $N$) for chiral extrapolations. 
While masses at each value of $\beta$ differ, 
by about 3\% in the largest case, 
the differences in the continuum limit do not exceed 1.5\%
of the results of Q$\chi$PT fits.

Compared to the results presented at Lattice'97\cite{ref:CPPACSlat97} 
where we employed a linear chiral extrapolation in $1/K$
(cubic for $N$ and quadratic for $\Lambda$), 
the nucleon and $\Delta$ masses have decreased by 4.5\% and 3.5\%, respectively.
Strange baryon masses with $m_K$ used as input have also decreased.
The shift, however, is within $1.5\sigma$ for all particles,
with either $m_K$ or $m_\phi$ as input.

In summary, we find that differences in chiral extrapolations and an increase 
of statistics at $\beta=6.47$ do not alter 
the conclusions we drew at the time of Lattice'97:  
With $m_\pi$, $m_\rho$ and $m_K$ used as input,
the meson hyperfine splitting and decuplet baryon mass splitting are 
too small compared to experiment, and so are the octet baryon masses. 
When we use $m_\phi$ instead of $m_K$ as input, the discrepancies for
baryon masses are reduced, but the meson hyperfine splitting remains smaller.
 
\section{Light quark masses}
\vspace{-0.2mm}
\begin{figure}[t]
\begin{center} \leavevmode
\epsfxsize=7.0cm \epsfbox{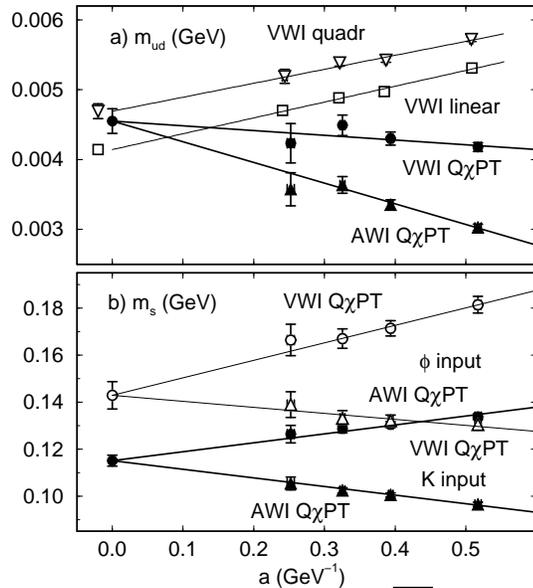}
\end{center}
\vspace{-15mm}
\caption{
Light quark masses in $\overline{\rm MS}$ scheme at $\mu=$ 2 GeV.}
\label{fig:QuarkMass}
\vspace{-7mm}
\end{figure}

The Q$\chi$PT fit to pseudo-scalar meson masses has a significant 
effect on light quark masses at finite $\beta$. 
Due to a negative curvature of the Q$\chi$PT formula, 
values of the averaged $u$ and $d$ quark mass  
defined with vector Ward identity $m_{ud}^{VWI}$
become smaller than those from polynomial 
chiral extrapolations as shown in Fig.~\ref{fig:QuarkMass}.  
The results extrapolated to the continuum limit, 
however, are consistent among various definitions. 
We adopt a combined fit to $m_q^{VWI}$ and $m_q^{AWI}$, both estimated 
with Q$\chi$PT fits, to calculate our final result. We obtain
$m_{u,d}$=4.6(2) MeV, and 
$m_s$=115(2) MeV ($m_K$ input) or 
143(6) MeV ($m_\phi$ input) in the ${\overline{\rm MS}}$ scheme 
at $\mu=$ 2~GeV.

\section{PS meson decay constants}
\vspace{-2mm}
We determine $f_\pi$ and $f_K$ from the local axial current, 
employing a quadratic polynomial and linear chiral extrapolation, respectively.
We obtain $f_\pi=$ 120(6) MeV and $f_K=$ 139(4) MeV, 
which are 10 and 15\% smaller than experiment, respectively. 
The ratio $f_K/f_\pi-1 =$ 0.156(29) is also smaller than experiment.

\section{Conclusions}
We have presented our final results on the quenched 
light hadron spectrum and related quantities. 
In the course of analyses 
we found that our data for light hadron masses are consistent with 
predictions of Q$\chi$PT.  The effect of Q$\chi$PT singularities is 
small, however, and the continuum results do not noticeably shift
from those obtained with polynomial chiral extrapolations. 
Our results show that the quenched light hadron spectrum 
clearly and systematically deviates from the experimental spectrum. 
The discrepancy is much larger than our statistical error
of 1\% for mesons and 2--3\% for baryons.

\vspace{2mm}
This work is supported in part by the Grants-in-Aid
of Ministry of Education
(Nos. 08640404, 09304029, 10640246, 10640248, 10740107).
GB, SE, and KN are JSPS Research Fellows. HPS is supported
by JSPS Research for Future Program.

\end{document}